\documentstyle[proceedings]{crckapb}

\def\apj#1{{\em Astrophys. J.} {\bf #1}}
\def\mn#1{{\em Mon. Not. R. astr. Soc.} {\bf #1}}
\def\aa#1{{\em Astron. Astrophys.} {\bf #1}}
\def\aas#1{{\em Astron. Astrophys. Suppl.} {\bf #1}}

\def\nat#1{{\em Nature,} {\bf #1}}
\def\apjs#1{{\em Astrophys. J. Suppl.} {\bf #1}}
\def\be{\begin{equation}}
\def\ee{\end{equation}}
\def\bea{\begin{eqnarray}}
\def\eea{\end{eqnarray}}
\def\etal{{\it et al.}\ }
\def\Mpc{$h_{100}^{-1}$~{\rm Mpc}}

\begin{opening}

\title{A 120~MPC SCALE IN  THE UNIVERSE}

\author{J. EINASTO}

\institute{Tartu Observatory, EE-2444 T\~oravere, Estonia}
\end{opening}

\runningtitle{120~MPC SCALE}

\begin{document}

\section{Introduction}

Galaxies are not distributed randomly but are concentrated within
elongated filamentary chains which consist of groups and clusters of
galaxies; the space between filaments is devoid of galaxies. Such
distribution can be called cellular~\cite{je78,zes82}: a cell is a
large low-density region surrounded by superclusters.  Examples are
the Northern Local Void surrounded by the Local, Coma and Hercules
superclusters~\cite{l95}, and the Bootes void~\cite{k81} surrounded by
the Hercules and Bootes superclusters.  Superclusters and voids form a
continuous network of alternating high- and low-density regions; the
mean diameter of voids between clusters in the supercluster-void
network is about 100~\Mpc~\cite{zes82}.

It is not clear whether superclusters and voids form a regular or
irregular network.  According to the classical paradigm of the
formation of large scale structure the distribution of density waves
is Gaussian on all scales, and thus the supercluster-void network
should have a random character. The observed network was formed by
density waves of a wavelength range corresponding to the scale of the
network. Therefore, it was a great surprise when it was found that the
distribution of high-density regions along pencil beams around the
northern and southern Galactic pole is fairly regular: high- and
low-density alternate with rather constant step of 128~\Mpc\
\cite{beks90}.  The regularity of the structure is so far well
established only in the direction of Galactic polar caps, while in
other directions the regularity is much less pronounced. In order to
find the degree of the global regularity of the supercluster-void
network 3-dimensional data of the distribution of high-density regions
are needed.  For this purpose Abell-ACO clusters of
galaxies \cite{abell,aco} can be used. These clusters cover the whole
celestial sphere outside the Milky Way zone of avoidance. Here I give
a summary of our principal results, a more detailed analysis is
published elsewhere.

\section{Distribution of clusters of galaxies}

The distribution of clusters of galaxies located in very rich
superclusters with at least 8 member-clusters is shown in Figure~1~
\cite{me94,me95,me97}, see also \cite{tully}. This Figure shows
clearly that high-density regions are separated from each other by a
fairly constant intervals of $\approx 120$~\Mpc; in other words, they
form a quasi-regular network of superclusters and voids.

\begin{figure}
\vspace{8cm}
\caption{Distribution of clusters in high-density regions in supergalactic
  coordinates (Einasto 1995). Only clusters in superclusters with at least 8
  members are plotted.  The supergalactic $Y=0$ plane coincides almost exactly
  with the Galactic equatorial plane; the Galactic zone of avoidance is marked
  by dashed lines.
} 
\includegraphics{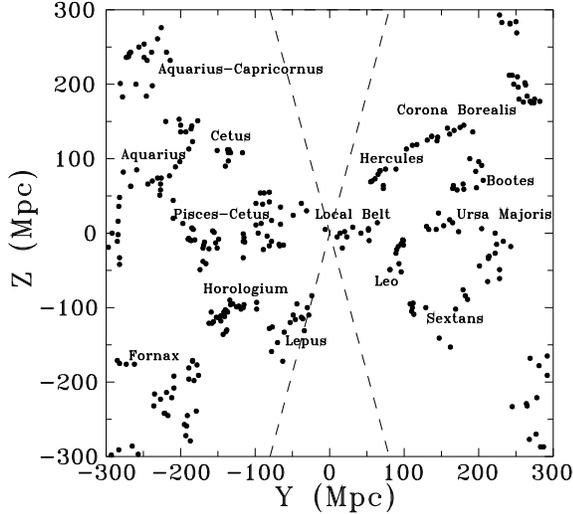}
\end{figure}

We supplemented this qualitative description by a quantitative
analysis, using the power spectrum of clusters of galaxies
\cite{einasto}.  On short wavelengths the spectrum can be approximated
by a power law with index $n=-1.8$. On wavenumber
$k_{0}=0.05$~$h_{100}~{\rm Mpc}^{-1}$ it has a sharp peak; this
wavenumber corresponds to the wavelength
$\lambda_0=2\pi/k_{0}=120$~\Mpc. On longer scales the error corridor
of the spectrum is large and here the spectrum is compatible with the
Harrison-Zeldovich spectrum of power index $n=1$.

The presence of a sharp maximum in the power spectrum of matter is the
main finding of our study of the distribution of clusters of galaxies.
Our result has found independent support from other data
\cite{landy,gaztanaga,peacock,retzlaff}.  Comparison with simple toy models
shows that a peaked power spectrum is possible only if high-density
regions form a quasiregular rectangular network
\cite{einasto1,einasto2}. In this case the correlation function of
objects located in high-density regions is oscillating. Evidence for
an oscillating cluster correlation function has been accumulating
already for some time \cite{kopylov,mo,eg93,fetisova}.

\section{Comparison with CMB data}

The angular power spectrum of the cosmic microwave background (CMB)
has been measured by a number of teams.  We compared the CMB spectrum
with optical data using three models for the power spectrum: (a) a
scale-free initial spectrum, (b) a double power law approximation to
the cluster spectrum, and (c) a spectrum based on the observed cluster
spectrum.  For a set of cosmological parameters, we calculate the
matter transfer function, and the matter and radiation power spectra
for all three models \cite{atrio}.  We assume the Universe has a flat
geometry. In calculations we used the package CMBFAST
\cite{cmbfast2}. To estimate the goodness of a particular set of
cosmological parameters we calculate the parameter $\chi^2$ for all
three principal models, see Figure~2.

\begin{figure}
\vspace*{13cm}
\caption{
  Goodness-of-fit contours of $\chi^2$ at 68\% and 95\% confidence
  level.  The $\chi^{2}$ statistics measures the deviation of the
  expected temperature anisotropy amplitude of a given model from the
  Saskatoon data.  The first row displays the results for the scale
  free model; the second row for the double power law model; and the
  lowest row for the cluster spectrum based model.  In the first
  column we plot models with varying Hubble constant and baryon
  fraction for a spectral index $n=1$ at large scales and no
  cosmological constant. In the middle column the same diagrams were
  repeated for $n=1.2$. Dashed lines indicate the nucleosynthesis
  bounds. The last column displays the results for
  models with different values of the 
  cosmological constant. On all these models,
  the age of the Universe was chosen to be $14$ Gyr.}
\includegraphics{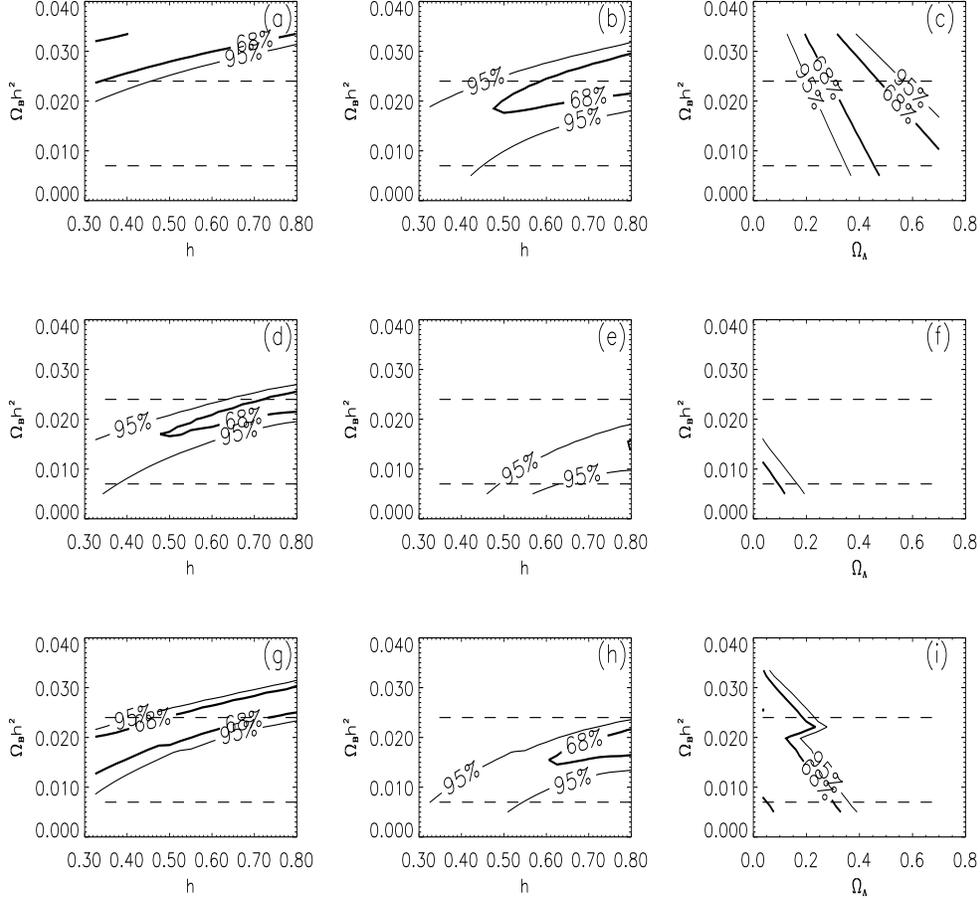}
\end{figure}

\begin{figure}
\vspace{6.5cm}
\caption{
  Comparison of matter power spectra and radiation temperature anisotropies
  with cluster and CMB data.  Dots with $1\sigma$ error bars give the
  observations: the measured cluster spectrum (Einasto \etal 1997a) in the
  left panel and the Saskatoon data on CMB temperature anisotropies
  (Netterfield \etal 1997) in the right panel. The scale-free model spectra
  (short-dashed lines) were computed using the following parameters: $h=0.6$,
  $\Omega_{b}=0.07$, $\Omega_{c}=0.23$, and $\Omega_{\Lambda}=0.7$.  
  The cluster spectrum (solid lines) was calculated
  using $h=0.6$, $\Omega_{b}=0.08$, $\Omega_{c}=0.92$, and
  $\Omega_{\Lambda}=0$; and the double power law models (long-dashed) using
  $h=0.6$, $\Omega_{b}=0.05$, $\Omega_{c}=0.95$, and $\Omega_{\Lambda}=0$.  To
  compare matter power spectra and observations we used a bias factor
  $b_{cl}\approx 3$ for cluster spectrum.} 
\includegraphics{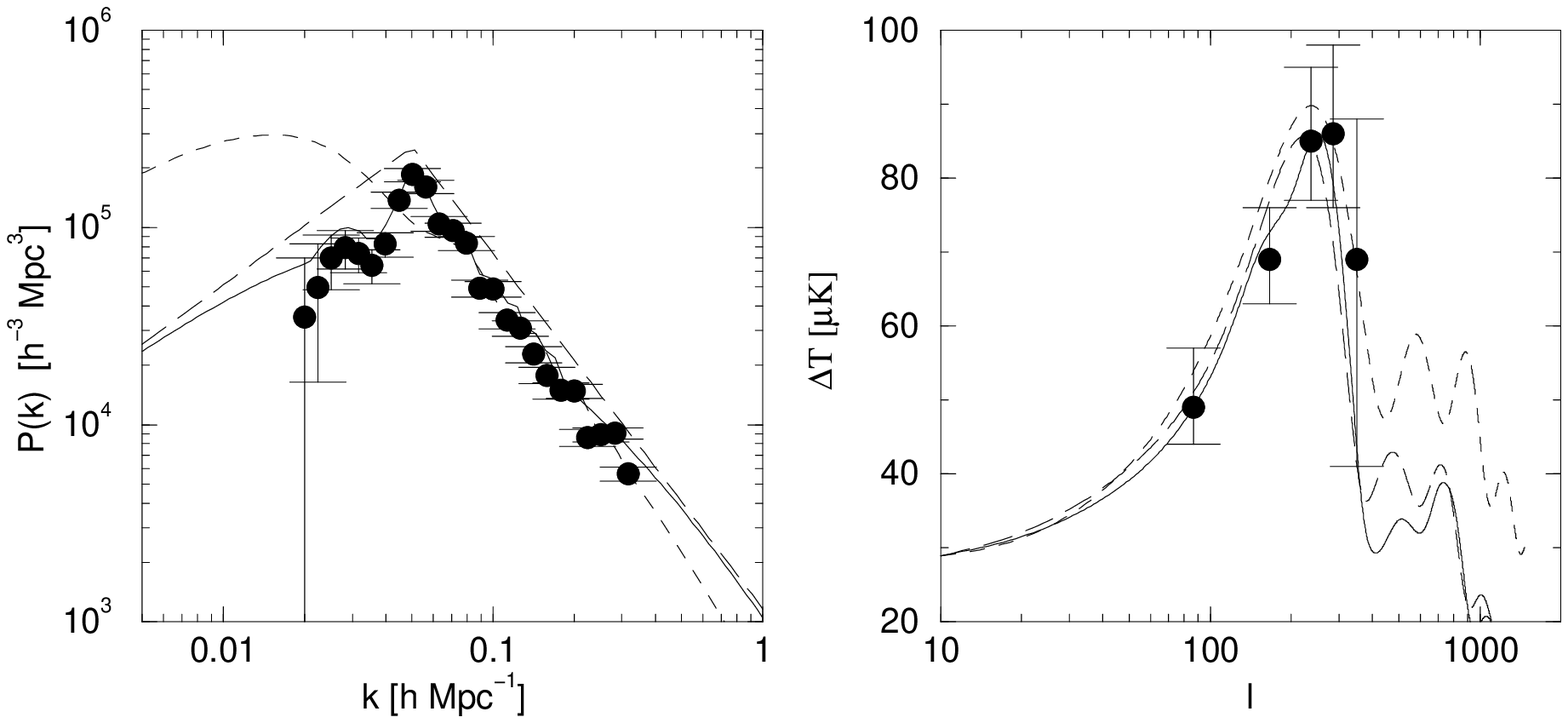}
\end{figure}

In Figure~3 we compare matter power spectra and temperature anisotropy
spectra for our three basic models with the data.  The cosmological
parameters were chosen to reproduce the CMB data~\cite{saskatoon1}. We
see that the temperature anisotropy spectra are very similar in the
range of multipoles observed in Saskatoon. In other words, the present
CMB data are not sufficient to discriminate between models.  The
matter power spectra are also similar on short wavelengths but on
medium and long scales they are very different. The scale-free model
with large cosmological constant has a broad maximum at large
wavenumber ($k\approx 0.01~h_{100}~{\rm Mpc^{-1}}$); the maximum of
the first acoustic oscillation occurs at $k\approx 0.1~h_{100}~{\rm
Mpc^{-1}}$ and is of rather small relative amplitude.  Both scales are
outside the allowed range of the observed spike in the cluster
spectrum: $k_{0}=0.052 \pm 0.005~h_{100}~{\rm Mpc^{-1}}$
\cite{einasto}.  No combination of cosmological parameters can
reproduce the spike at $k=k_{0}$: the existence of a broad maximum is
an intrinsic property of all scale-free models.  Thus the observed
spike is not related to acoustic oscillations in the baryon--photon
plasma as assumed by Szalay (1997) but must have a different origin.
On the other hand, the cluster and double power law spectra fit the
observed cluster spectrum by construction and reproduce the CMB data,
i.e they fit equally well both datasets.

Thus the comparison of optical and CMB data brings us to the
conclusion that CMB data are not in conflict with the presence of a
spike in the matter power spectrum, and that the present combined
cluster and CMB data favour models with a built-in scale in the {\it
initial} spectrum.  We repeat that a regular supercluster-void network
can be formed only by a power spectrum with a sharp maximum
\cite{einasto1}. As noted by Szalay (1997) correlated phases are also
crucial to form a regular network of high- and low-density regions.

Double inflation models provide a possible scenario where the formation of a
spike could have taken place. One version of a double inflation model is
suggested by Starobinsky (1992).  This model produces a spectrum
rather similar to the initial spectrum found from data~\cite{atrio}.  The
study of the distribution of matter on large scales is of crucial importance
since it could provide a direct test of more complicated models of inflation.

\vspace{0.5cm}

I thank H. Andernach, F. Atrio-Barandela, M. Einasto, S.  Gottl\"ober,
V.  M\"uller, A. Starobinsky, E. Tago and D. Tucker for fruitful
collaboration and permission to use our joint results in this review
article, and A. Szalay for discussion.  This study was supported by
the Estonian Science Foundation.

\end{document}